\documentclass[a4paper,fleqn]{cas-dc}

\usepackage[numbers,sort&compress]{natbib}
\usepackage{color}
\usepackage{float}
\usepackage{algorithm}
\usepackage{cleveref}
\def\tsc#1{\csdef{#1}{\textsc{\lowercase{#1}}\xspace}}
\tsc{WGM}
\tsc{QE}
\tsc{EP}
\tsc{PMS}
\tsc{BEC}
\tsc{DE}
\begin{document}
\let\WriteBookmarks\relax
\def\floatpagepagefraction{1}
\def\textpagefraction{.001}
\shorttitle{}
\shortauthors{Fei Jiao et~al.}

\title [mode = title]{The Layer-inserting Growth of Antiferromagnetic Topological Insulator MnBi$_2$Te$_4$ Based on Symmetry and Its X-ray Photoelectron Spectroscopy}

\author[1,4]{Fei Jiao}
\address[1]{School of Materials Science and Engineering, Shandong University of Technology, 266 Xincun Xi Road, Zibo, 255000, China}
\author[4]{Jingfeng Wang}[style=chinese]
\author[1,4]{Xianyu Wang}[style=chinese]
\author[4]{Qingyin Tian}[style=chinese]
\author[3]{Meixia Chang}[style=chinese]
\author[4]{Lingbo Cai}[style=chinese]
\author[4]{Shu Zhu}[style=chinese]
\author[4]{Di Zhang}[style=chinese]
\author[4]{Qing Lu}[style=chinese]
\author[4]{Cao Wang}[style=chinese]
\author[4]{Shugang Tan}[style=chinese]
\author[2]{Yunlong Li}[style=chinese]
\author[2]{Jiayuan Hu}[style=chinese]
\author[2,4]{Qiang Jing}[type=editor,]
\cormark[1]
\ead{jingqiang@sdut.edu.cn}
\address[2]{ Key Laboratory of Artificial Structures and Quantum Control (Ministry of Education),
School of Physics and Astronomy, Shanghai Jiao Tong University, Shanghai 200240, China}
\author
[1]
{Bo Liu}
\cormark[2]
\ead{liub@sdut.edu.cn}
\author[2]{Dong Qian}[style=chinese]

\address[3]{ Key Laboratory for Magnetism and Magnetic Materials of Ministry of Education,Lanzhou University}
\address[4]{ Laboratory of Functional Molecules and Materials, School of Physics and Optoelectronic Engineering,Shandong University of Technology, 266 Xincun Xi Road, Zibo, 255000, Chinas }

\cortext[cor1]{Corresponding author}
\cortext[cor2]{Corresponding author}

\credit{Data curation, Writing - Original draft preparation}

\begin{abstract}
The antiferromagnetic topological insulator has attracted lots of attention recently, as its intrinsic magnetism and topological property makes it a potential material to realize the quantum anomalous Hall effect (QAHE) at relative high temperature. Until now, only MnBi$_2$Te$_4$ is predicted and grown successfully. The other  MB$_2$T$_4$-family materials predicted (MB$_2$T$_4$:M=transition-metal or rare-earth element, B=Bi or Sb, T=Te, Se, or S) with not only antiferromagnetic topological property but also rich and exotic topological quantum states and dynamically stable (or metastable) structure have not been realized on experiment completely. Here, MnBi$_2$Te$_4$ single crystals have been grown successfully and tested. It shows typical antiferromagnetic character, with Neel temperature of 24.5K and a spin-flop transition at H$\thickapprox$35000 Oe, 1.8K. After obtaining MnBi$_2$Te$_4$ single crystals, we have tried to synthesize the other members of  MB$_2$T$_4$-family materials, but things are not going so well. Then it inspires us to discuss the growth mechanism of MnBi$_2$Te$_4$. The growth mode may be the layer-inserting growth mode based on symmetry, which is supported by our X-ray photoelectron spectroscopy (XPS) measurement. The XPS measurement combing with the $Ar^+$ ion sputtering is done to investigate the chemical state of MnBi$_2$Te$_4$. Binding energies (BE) of the MnBi$_2$Te$_4$-related contributions to Mn2p and Te3d spectra agree well with those of inserting material $\alpha$-MnTe. Rising intensity of the Mn2p satellite for divalent Mn (bound to chalcogen) with atomic number of ligand (from MnO to MnBi$_2$Te$_4$) has been observed, thus suggesting classification of MnBi$_2$Te$_4$ as the charge-transfer compound. Understanding the growth mode of MnBi$_2$Te$_4$ can help us to grow the other members of  MB$_2$T$_4$-family materials.
\end{abstract}

\begin{graphicalabstract}
\includegraphics[width=19cm]{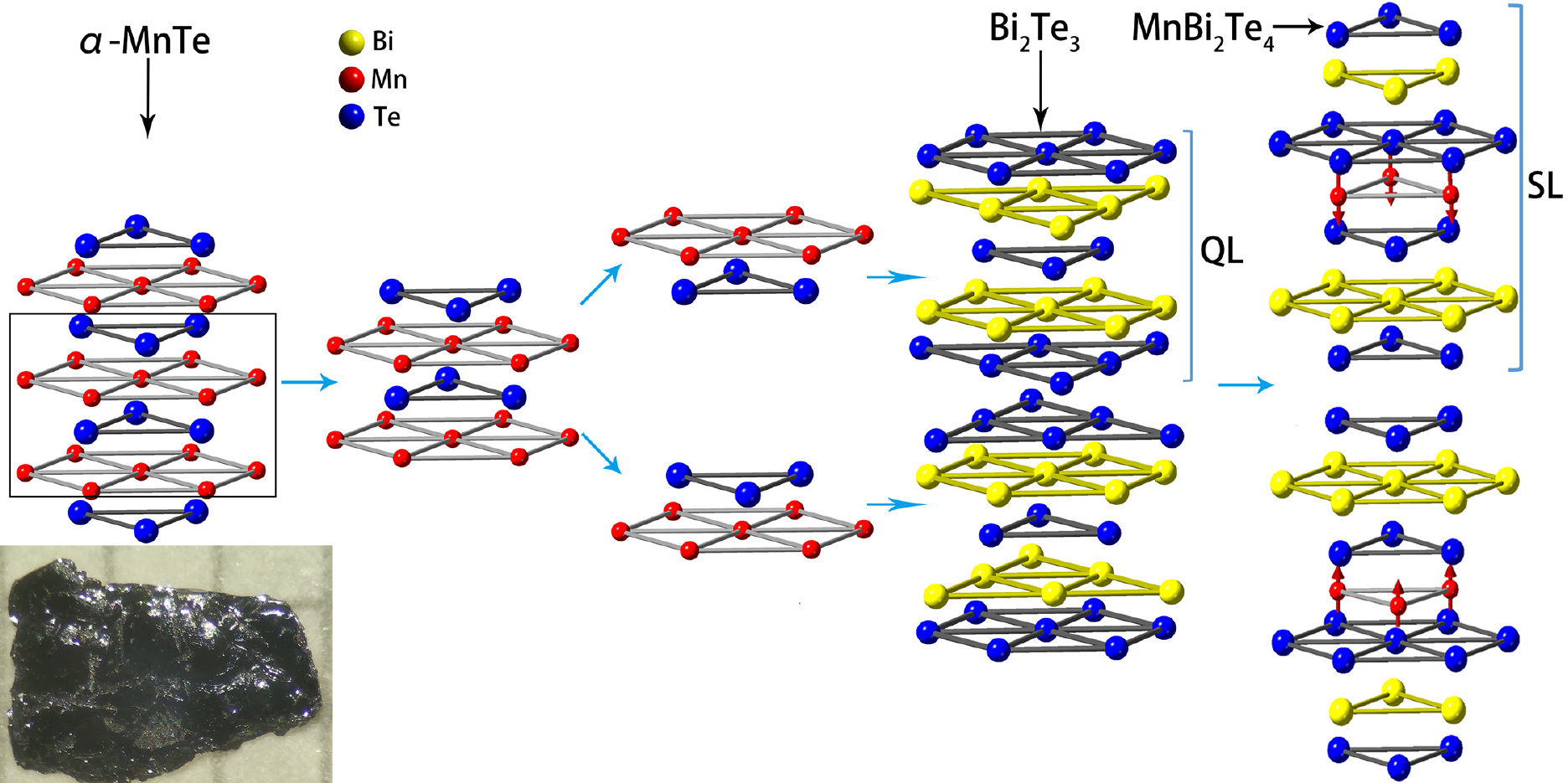}
\end{graphicalabstract}

\begin{highlights}
\item A possible layer-inserting growth mode for antiferromagnetic topological insulator MnBi$_2$Te$_4$ is proposed.
\item   The chemical state of MnBi$_2$Te$_4$ has been investigated by XPS.
\item   The chemical states of Mn and Te in MnBi$_2$Te$_4$ agree well with those of
$\alpha$-MnTe .
\end{highlights}

\maketitle

\section{Introduction}

The milestone discoveries of integer quantum Hall effect (QHE) and fractional QHE made the QHE one of the most important fields in modern condensed mater physics. The dissipationless chiral edge states of the QHE regime can be used in low power consumption, high speed electronic devices. The unique chiral edge states responsible for the QHE originate from the magnetically induced Landau levels\cite{1}.
However, well-defined Landau levels are only possible in high-mobility samples under strong external magnetic fields. The demanding requirements prevent the QHE from being widely applied in industry. Therefore, it is highly desirable to achieve the QHE without the need of a strong magnetic field and an extraordinarily high mobility sample. The anomalous Hall effect (AHE) in a ferromagnet can be induced by spontaneous magnetization without needing an external magnetic field \cite{2}. Then, a quantized version of the AHE,  the quantum anomalous Hall effect (QAHE), representing the realization of the QHE in the zero magnetic field has come to the forefront \cite{2,3}.When QAHE is realized, one can observe a plateau of Hall conductance ($\rho_{xy}$) of  $e^2$/h and a vanishing longitudinal conductance ($\rho_{xx}$) even at zero magnetic field (Fig.1)\cite{4}. Then, the realization of the QAHE may lead to the development of low-power-consumption electronics. There are two prerequisites for the observation of the QAHE : the formation of an exchange gap induced by the coupling with the magnetization in or next to the topological insulator (TI) and the precise tuning of E$_F$ into the gap\cite{5}. In detail, to realize QAHE requires four conditions: (1) the system must be 2D; (2) insulating bulk; (3) ferromagnetic ordering; and (4) a non-zero Chern number\cite{6}. Then topological materials holding magnetism can satisfy the conditions \cite{7,8,9}. Until now, QAHE state has only been realized in magnetically doped Cr- and/or V-doped (Bi,Sb)$_2$Te$_3$ topological insulator films system, below $\sim$ 1K \cite{4,10}. However, such a low critical temperature severely constrains both the exploration of fundamental physics and technological applications based on this exotic phenomenon\cite{11}.  Therefore, seeking new materials to realize QAHE at higher temperature has become a major research direction in the field of topological quantum materials.
{One approach is to try to dop the other 3d elements such Mn, Fe and Co into the TI. The opening of the gap has been realized in Fe or Mn-doped Bi$_2$Se$_3$\cite{60,61}. For Co doped TI, ferromagnetic behaviour and antiferromagnetic behaviour have been observed in Bi$_{1-x}$Co$_x$Se$_3$ \cite{72,73} and Sb$_{2-x}$Co$_x$Te$_3$ \cite{74}, respectively. Another approach is to use the magnetic proximity effects (MPEs) to induce magnetic order in a TI by coupling it to an adjacent magnetically ordered material.  Lots of excellent works have been done such as the reports  on EuS/TI \cite{62,63,64}, Y$_3$Fe$_5$O$_{12}$/TI \cite{65, 66, 67}, CrSb/TI\cite{68},Tm$_3$Fe$_5$O$_{12}$/TI \cite{69} and CrSe/TI\cite{70}.} Another promising approach is to realize high-temperature QAHE insulator in thin films of intrinsic ferromagnetic (FM) or antiferromagnetic (AFM) TI materials \cite{12}. Nevertheless, despite much theoretical and experimental efforts devoted, there is little progress until the recent discovery of an intrinsic AFM TI MnBi$_2$Te$_4$ \cite{13,14,15,16,17,18}. {The growth of MnBi$_2$Te$_4$ was first reported by  Lee et al. and it was treated as a kind of thermoelectric material at that time\cite{32}. Then, lots of theoretical and experimental work were done to treat MnBi$_2$Te$_4$ as a kind of antiferromagnetic topological \\insulator\cite{13,14,15,16,17,18}. On experiment, MnBi$_2$Te$_4$ could be synthesized through several approaches, but it is  not so easy to get  the high purity crystalline samples with perfect  antiferromagnetic order\cite{18,22}. Then to understand its growth mechanism become an urgent necessity.}In another aspect, it is proposed that MnBi$_2$Te$_4$ is just one member of  MB$_2$T$_4$-family materials ( MB$_2$T$_4$: M = transition-metal or rare-earth element, B = Bi or Sb, T = Te, Se, or S) \cite{14,19, 20}, which have rich topological quantum states. Among them, MnBi$_2$Te$_4$ has been grown successfully and tested, but the other materials of this family with dynamically stable structure have not been realized completely until now. Then,understanding the growth mechanism can help researcher to get the single crystals of the other family members too. {Recently, Li et al. has reported the growth of high quality MnBi$_2$Te$_4$ with perfect  antiferromagnetic order\cite{18}. Based on the result of high-resolution high-angle annular dark field scanning transmission electron microscopy (HAADF-STEM), they proposed the growth mechanism of MnBi$_2$Te$_4$ was the intercalation of MnTe into Bi$_2$Te$_3$\cite{33}. Be synchronous with their work, we have grown high quality MnBi$_2$Te$_4$ single crystal with perfect  antiferromagnetic order too. Our experiences in the synthesis of layered iron-based superconductor ThFeAsN, ThCoAsN, ThMnAsN et.al.\cite{34, 35, 36}with ThN layer as the block layer inspire us to consider that the growth mode of MnBi$_2$Te$_4$ may be the layer-inserting growth mode. To check it, the chemical states of MnBi$_2$Te$_4$ and $\alpha$-MnTe were investigated by
XPS combing with the $Ar^+$ ion sputtering. However, the measurement result supports our guess.  Understanding the growth mode of MnBi$_2$Te$_4$ can help us to grow the other members of  MB$_2$T$_4$-family materials.  Recently, there are new research progress on the intrinsic magnetic topological insulator and QAHE too. There are several reports on the van-der-Waals-gap inserting growth of (MnBi$_2$Te$_4$)(Bi$_2$Te$_3$)$_m$ i.e. inserting Bi$_2$Te$_3$ to the van der Waals gap of MnBi$_2$Te$_4$ \cite{37,38,39,40}. With the increasing of m, the interlayer antiferromagnetic exchange coupling of \\(MnBi$_2$Te$_4$)(Bi$_2$Te$_3$)$_m$ becomes gradually weakened, as the separation of magnetic layers increases. With m=3, it shows long-range ferromagnetic order below 10.5 K, with the easy axis along the c axis\cite{40}. At the same time, the zero-field QAHE has been realized in pure MnBi$_2$Te$_4$ at 1.4K by Deng et. al., recently\cite{41}.
}
\begin{figure}
	\centering
		\includegraphics[width=8.5cm]{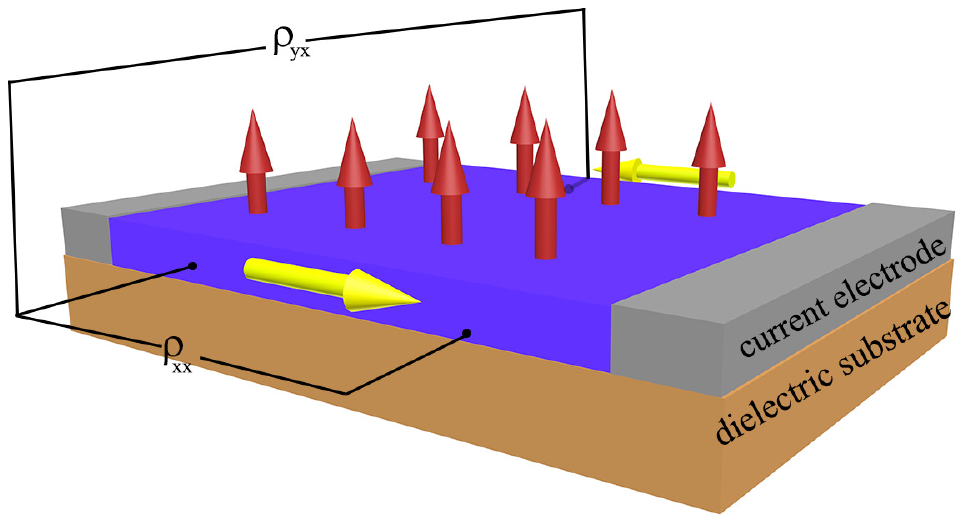}
	\caption{A schematic diagram depicting the principle of  QAHE based on a topological insulator thin film owing ferromagnetism. The red arrows have directed the magnetization direction (M). The yellow arrows denote the currents, which are from the conducting edge channel of the device.}
	\label{FIG:1}
\end{figure}
\begin{figure*}
	\centering
		\includegraphics[width=19cm]{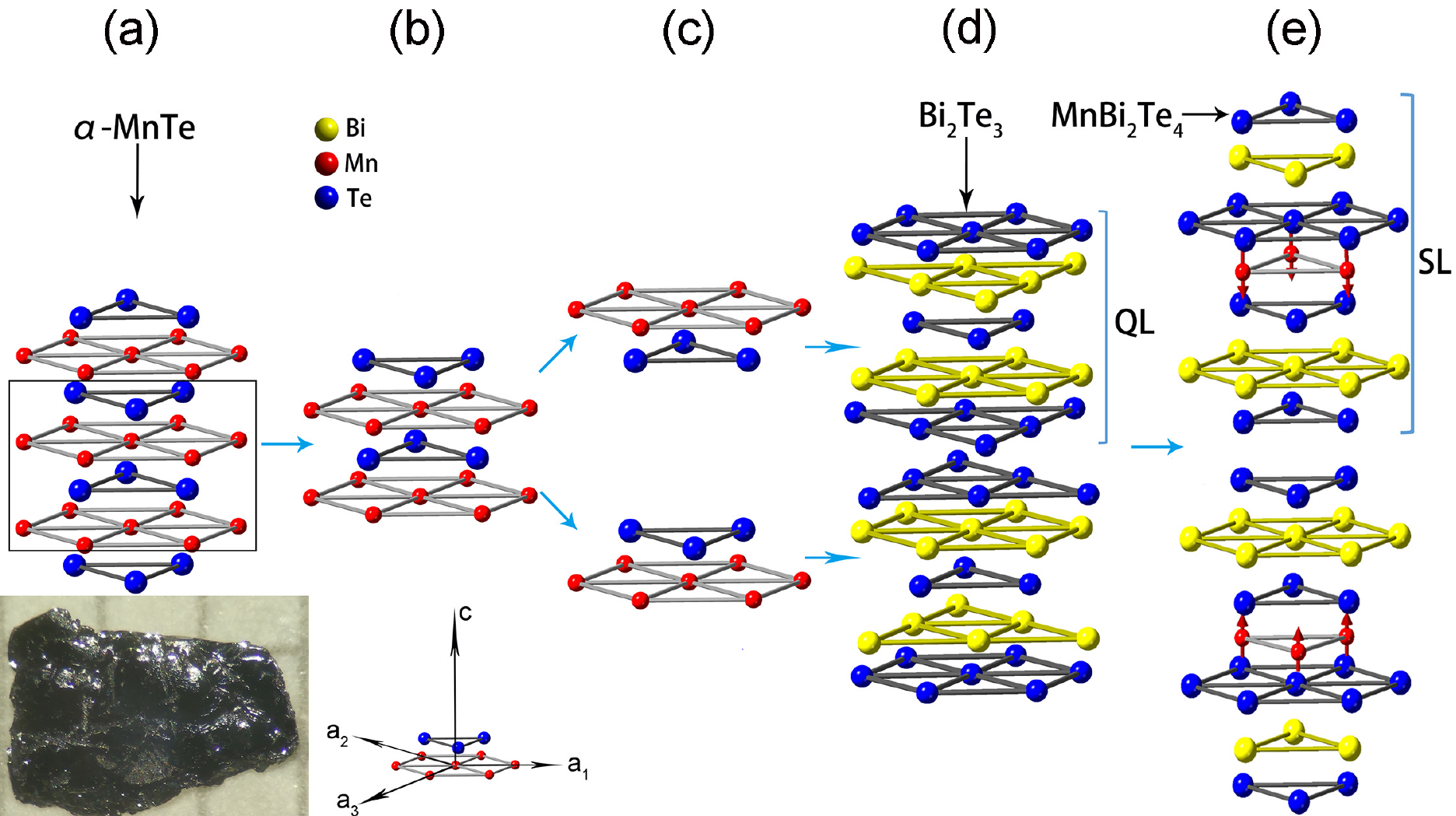}
	\caption{The illustration of layer-inserting growth process of MnBi$_2$Te$_4$. MnBi$_2$Te$_4$ includes seven atoms in a unit cell, forming a Te-Bi-Te-Mn-Te-Bi-Te septuple layer (SL).The inserted part of the figure is the single crystal of MnBi$_2$Te$_4$. It is about 3×1.5$mm^2$ in dimension.}
	\label{FIG:2}
\end{figure*}
\section{Experimental Section}
\subsection{Growth of MnBi$_2$Te$_4$ Single Crystals}
\par
Bi powder (99.99\%), Te powder (99.999\%) and Mn powder (99.9\%) were all bought from Aladdin Company of China.
$\alpha$-MnTe polycrystalline samples were prepared by mixing the powder components of Mn and Te in an atomic ratio 1:0.97. Then the mixture was placed in an Al$_2$O$_3$ crucible and sealed in an evacuated quartz tube. The quartz tube was initially heated up to 700$^{\circ}$C at a rate of 2$^{\circ}$C/h, holding at this temperature for one day followed by water quenching. The growth process of Bi$_2$Te$_3$ is as following: Bi powder and Te powder were mixed according to the stoichiometric ratio. Then the mixture was placed in an Al$_2$O$_3$ crucible and sealed in an evacuated quartz tube. The quartz tube was initially heated up to 850$^{\circ}$C in 5 hours and then kept at this temperature for about 20 hours. Next,  the temperature was decreased to 550$^{\circ}$C in 48 hours. After that it was kept at 550$^{\circ}$C for 5 days then the heating program was closed. After that the Bi$_2$Te$_3$ singe crystal was gotten and then was smashed into powder to be utilized as precursor.
High quality single crystals of MnBi$_2$Te$_4$ were synthesized using the flux method. The raw materials of $\alpha$-MnTe and Bi$_2$Te$_3$ were mixed in the molar ratio of 1: 5.85 in an Al$_2$O$_3$ crucible, which were sealed inside a quartz ampule. The ampule was put into a furnace and heated to 950$^{\circ}$C over a period of one day. After maintaining it at 950$^{\circ}$C for 12h, the ampule was cooled to 580$^{\circ}$C at a rate of 10$^{\circ}$C/h. Large-sized MnBi$_2$Te$_4$ crystals were obtained after centrifuge in order to remove the excess Bi$_2$Te$_3$ flux.\par
\subsection{Physical Properties Measurements}
XPS were performed on Axis Ultra of Kratos Analytical with AlK$\alpha$ (h$\nu$=1486.6 eV) as the monochromatic X-ray radiation source. The electron analyzer pass energy was fixed at 30eV. The pressure in the spectrometer chamber was maintained at about 8×$10^{-10}Pa$. The binding energy (BE) scale was referenced to the energy of the C1s peak of adventitious carbon, E$_B$=284.8 eV. All peaks had been fitted using a Shirley background and Voigt (mixed Lorentzian-Gaussian) line shapes.
{In order to remove the surface contamination and oxidation layer, then to observe the  intrinsic chemical states of $\alpha$-MnTe and MnBi$_2$Te$_4$, sputtering procedure was done by utilizing the Argon ion gun. For both bulk single crystal MnBi$_2$Te$_4$ and polycrystal $\alpha$-MnTe, about 200nm-thick-film were removed within 180s. For the Argon ion gun, the estimated sputtering rate is about 1.1nm/sec and the ion energy is 1000eV.}The analysis software CasaXPS is used to fit the peaks. The magnetization measurements were performed in a Quantum Design superconducting quantum interference device vibrating sample magnetometer system (SQUID-VSM).

\section{Results and Discussion}

\subsection{Crystal Structure and XRD Result}

Bi$_2$Te$_3$ with rhombohedral crystal structure (trigonal phase) belongs to the space group of $D^5_{3d}$ $(R\bar{3}m$), with five-atom layers arranging along the c-direction in one unit cell forming a Te-Bi-Te-Bi-Te quintuple layer. The coupling is strong between two atomic layers within one quintuple layer but much weaker, predominantly of the van der Waals type, between two quintuple layers \cite{21}. Sharing the isostructure with Bi$_2$Te$_3$, MnBi$_2$Te$_4$ crystalizes in a rhombohedral layered structure belonging to $R\bar{3}m$ space group too. Similar with Bi$_2$Te$_3$, each layer of a unit cell has a triangular lattice with atoms ABC stacked along the c axis \cite{12,22}. Slightly differently, monolayer MnBi$_2$Te$_4$ includes seven atoms in a unit cell, forming a Te-Bi-Te-Mn-Te-Bi-Te septuple layer (SL), which can be viewed as intercalating a Mn-Te bilayer into the center of a Bi$_2$Te$_3$ quintuple layer (QL) (Fig.2).
\begin{figure*}
\centering
		\includegraphics[width=14cm]{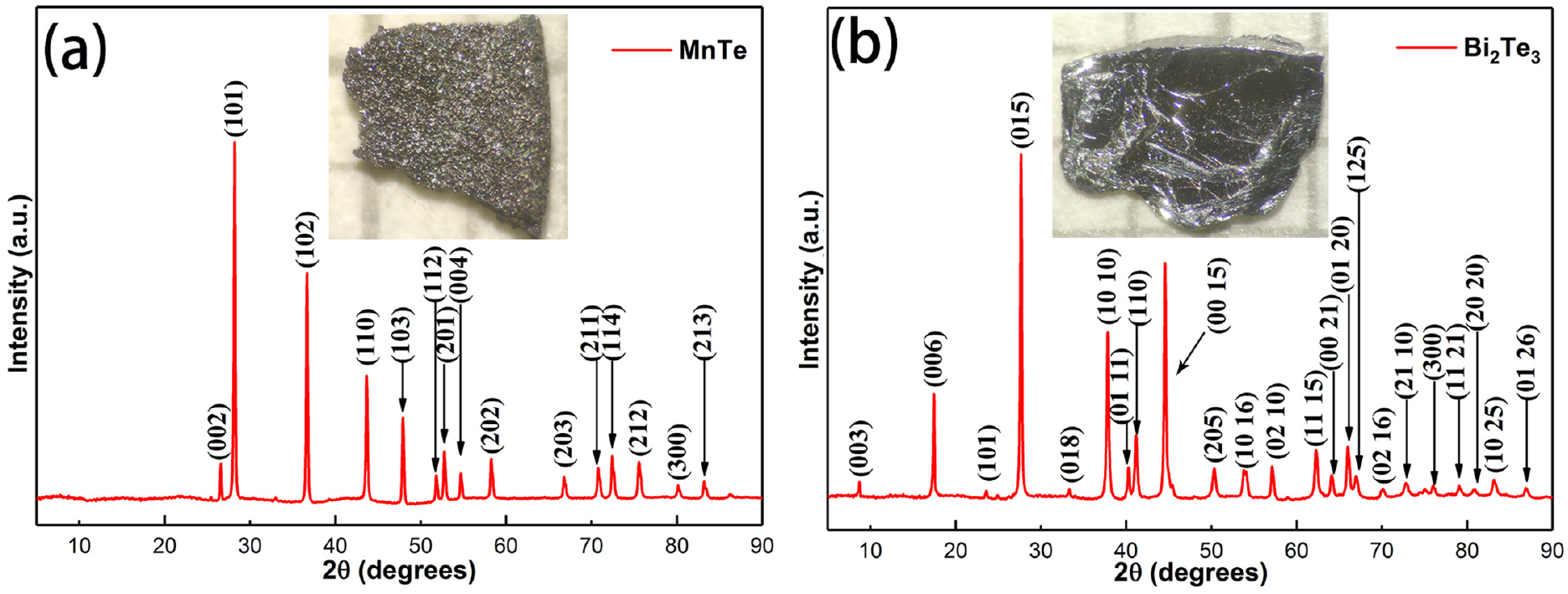}
	\caption{(a) and (b) are the powder XRD pattern of $\alpha$-MnTe and Bi$_2$Te$_3$, respectively. The inserted part of (a) and (b) are the crystal morphology of $\alpha$-phase MnTe polycrystals and Bi$_2$Te$_3$ single crystal, respectively.}
	\label{FIG:3}
\end{figure*}
As shown in the experimental section, the MnBi$_2$Te$_4$ is synthesized by self-flux method, through mixing $\alpha$-MnTe and Bi$_2$Te$_3$ together, using Bi$_2$Te$_3$ as the flux agent. After being heated to 950$^{\circ}$C and held for one day, the mixture are cooled slowly to 580$^{\circ}$C. During the crystal growth process, the MnTe layers inserts into the Bi$_2$Te$_3$ layers and then MnBi$_2$Te$_4$ single crystal are gotten. Here, we can use some theories summarized for layered iron-based superconductors to help us to understand the layer-inserting growth of MnBi$_2$Te$_4$\cite{23}. It may follow such key points: I) The inserting layers (MnTe bilayer) and the inserted layers (Bi$_2$Te$_3$) share the similar crystal structure. Bi$_2$Te$_3$ with trigonal phase belongs to $R\bar{3}m$(166) space group, while $\alpha$-MnTe with hexagonal phase belongs to P6$_3$/mmc (194) space group, which is stable in bulk form. As shown by Fig.2 that, both unit cells of $\alpha$-MnTe and Bi$_2$Te$_3$ have layered structure, with the trigonal layer and hexagonal layer alternately stacking along the c axis, which makes the insertion of -Mn-Te- possible. However, it is predicted that rhombohedral MnBi$_2$Se$_4$ ($R\bar{3}m$(166) space group) is also a kind of antiferromagnetic topological insulator\cite{19}.We have tried several times to get the bulk rhombohedral structure MnBi$_2$Se$_4$ by the similar inserting method, but it is not successful. What we have done is that, we use $\alpha$-MnSe of NaCl structure ($Fm\bar{3}m$ space group of cubic system and the stable phase of manganese selenium compounds) as the inserting layer material to try to insert it into the rhombohedral Bi$_2$Se$_3$. Then it can be explained from the symmetry perspective: inserting $\alpha$-MnSe with NaCl structure into rhombohedral (trigonal phase) Bi$_2$Se$_3$, just like inserting a square into a stack of triangle (or regular hexagon), which is not permitted by the symmetry and not stable too.We think the symmetry is the major obstacle to hinder the formation of rhombohedral MnBi$_2$Se$_4$ which is a potential antiferromagnetic topological insulator. II) Lattice mismatch between $\alpha$-MnTe and Bi$_2$Te$_3$ is acceptable. According to the powder XRD pattern shown by Fig.3, we have refined lattice parameters of $\alpha$-MnTe which are a=4.1477 {\AA}, c=6.7123 {\AA}, while for that of Bi$_2$Te$_3$ are a=4.384 {\AA}, c=30.488 {\AA}, respectively. As the lattice parameters mainly influencing the inserting process is from a axis,  the lattice mismatch between $\alpha$-MnTe and Bi$_2$Te$_3$ in a axis is calculated, which is only 3.4\%  ($\frac{a-{a_0}}{a}\times100\%$). \uppercase\expandafter{\romannumeral3}) Self-stability of $\alpha$-MnTe. The bulk $\alpha$-phase of MnTe is  stable at room temperature and atmospheric pressure. \uppercase\expandafter{\romannumeral4}) There exists charges transfer between the inserting layers (MnTe) and the inserted layers (Bi$_2$Te$_3$).
As shown by Fig.2 (c), (d) and (e) that, during the inserting growth process bilayer -Mn-Te- of $\alpha$-MnTe would insert into the gap between the third layer (Te atoms layer) and fourth layer (Bi atoms layer) of the unit quintuple layer of Bi$_2$Te$_3$. Therefore, Mn atoms of $\alpha$-MnTe would bond with Te atoms of Bi$_2$Te$_3$, then Mn atoms would transfer electrons to Te; Te atoms of $\alpha$-MnTe would bond with Bi atoms of Bi$_2$Te$_3$, then Bi atoms would transfer electrons to Te. The charges transfer can make the MnBi$_2$Te$_4$ system stable. In summary, the growth mode of MnBi$_2$Te$_4$ has typical character of inserting layer mode.\par
\subsection{Magnetic and Transport Properties }
\begin{figure*}
	\centering
		\includegraphics[width=14cm]{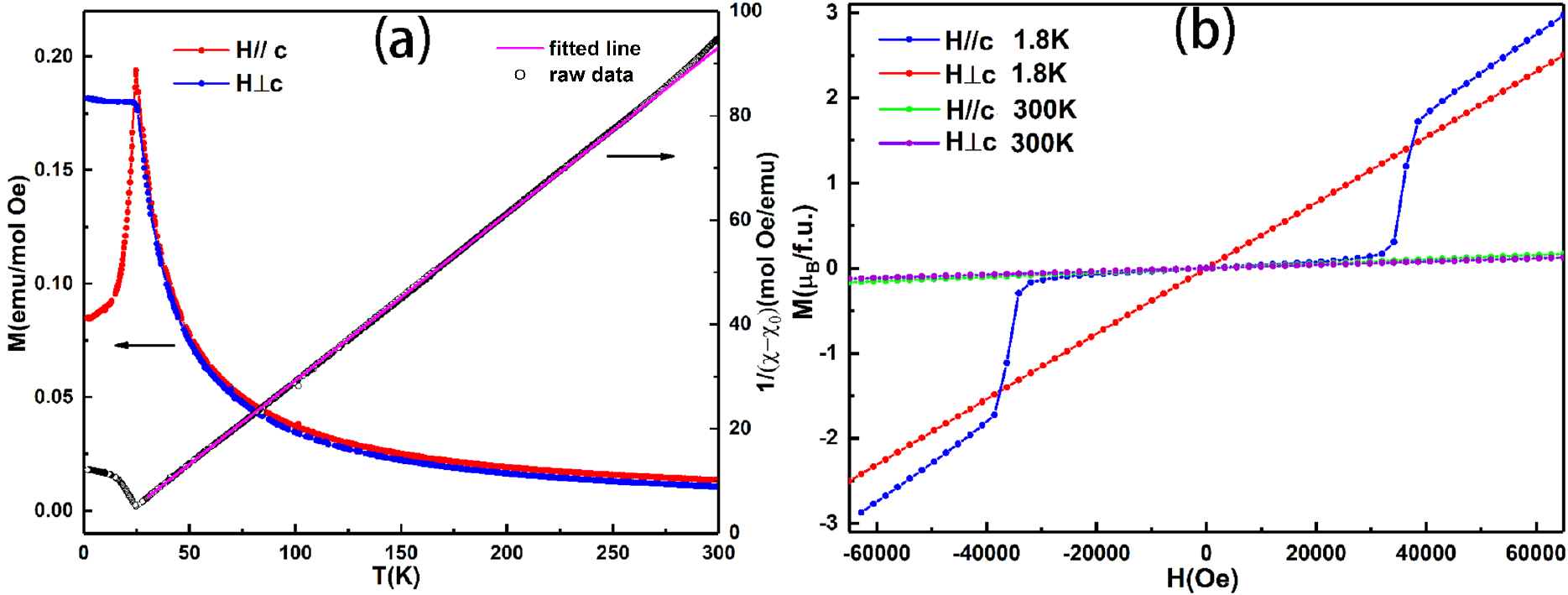}
	\caption{(a) The left axis of ordinates shows the magnetic susceptibility of MnBi$_2$Te$_4$ as a function of temperature measured in an external magnetic field of H=10000Oe in zero-field-cooled (zfc) mode. The right axis of ordinates shows the temperature dependent reciprocal susceptibility of H//c. The pink line is a fitted line according to modified Curie-Weiss law, fitted in the range of 30 K to 300 K ($\chi_0$=0.0029 emu/mol Oe). (b) Field-dependent magnetization curves for  two directions measured at 1.8K and 300 K.  }
	\label{FIG:4}
\end{figure*}
\begin{figure}
	\centering
		\includegraphics[width=8.5cm]{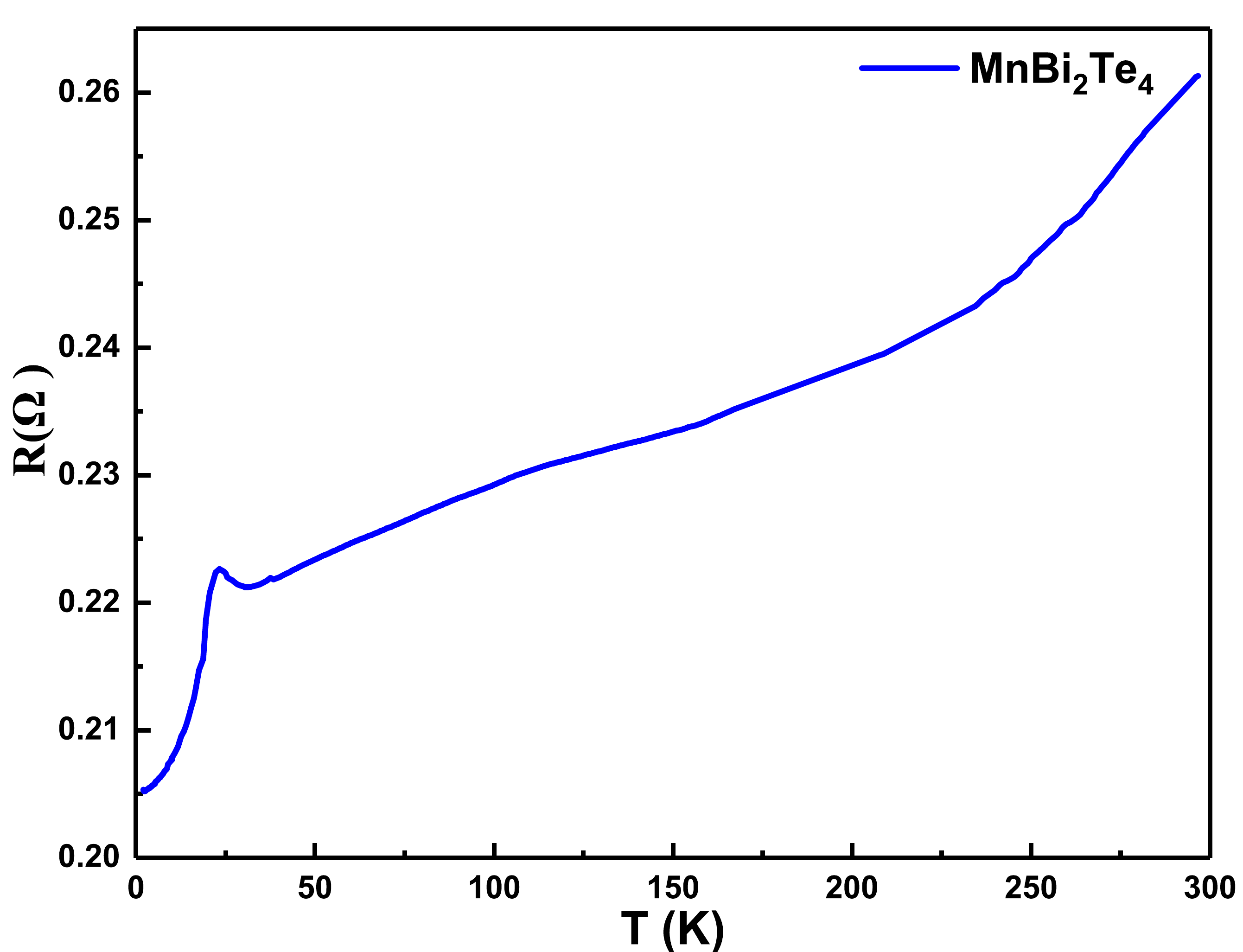}
	\caption{Temperature dependence of resistance of MnBi$_2$Te$_4$ single crystal measured from 2K to room temperature.}
	\label{FIG:5}
\end{figure}

Fig.4 (a) shows the temperature dependent magnetic susceptibility of MnBi$_2$Te$_4$ single crystal measured at 10000Oe. {It could be observed that MnBi$_2$Te$_4$ shows antiferromagnetic behaviour with T$_N$= 24.5K}. Meanwhile, a strongly anisotropic magnetic susceptibility is observed. The magnetic susceptibility  of H//c displays much steeper transition below 24.5K, which means the c axis is the easy axis. A modified Curie-Weiss law $\chi$(T)=$\chi_0$+C/(T-$\Theta_{cw}$) is used to fit the paramagnetic regime above $T_N$  in the range of 30 K to 300 K. Here, $\chi_0$ means a temperature-independent magnetic susceptibility, which contains two parts: one is diamagnetism closed electron shells; the other one is the Pauli paramagnetism due to the metallicity of MnBi$_2$Te$_4$. The temperature-dependent Curie-Weiss susceptibility C/(T-$\Theta_{cw}$) is mainly from localized Mn moments. The fitted effective paramagnetic moment with the value of 5.0$\mu_B$ roughly agrees with the high-spin configuration of $Mn^{2+}$ (S = 5/2)\cite{15}. Meanwhile, the Curie-Weiss temperature $\Theta_{cw}$=3.1K strongly depends on the fitted $\chi_0$ contribution. Below $T_N$, a spin-flop transition at H$\thickapprox$ 35000 Oe is observed for H//c (Fig. 4b) in the M-H curve, which is in line with the M-T result suggesting that c axis is the easy axis. {To check its magnetic properties further, field dependence of magnetization of MnBi$_2$Te$_4$ single crystal has been measured at different temperatures under the magnetic field of H//c and H$\perp$c. Its temperature dependence of magnetic susceptibility has been measured under magnetic field too(see Fig.S1 of the Supplementary materials). As shown by Fig.S1 (a) that, for H//c, above T$_N$, the magnetization displays a linear dependence, indicating the paramagnetic state; below T$_N$, all the curves show spin-flop transition. As shown by Fig.S1 (c) that, the antiferromagnetic transition tends to become weaker and finally disappears with the increase of the applied out-of-plane magnetic field, due to the suppression and spin-flop of interlayer antiferromagnetic coupling caused by the external out-of-plane magnetic field\cite{16,18}.}
The magnetic measurement result shows that the MnBi$_2$Te$_4$ single crystal owns good crystal quality and well antiferromagnetic property. {As shown by Fig.5, complementary to the magnetic measurements, the temperature dependence of the in-plane electrical resistance R(T) of a bulk MnBi$_2$Te$_4$ single crystal was measured from room temperature down to T = 2.0 K. The resistance  decreases with lowering the temperature, which is the metallic behaviour. The R-T curve exhibits a distinct peaklike feature centered at 24.0 K. The local increase indicates an enhanced electron scattering in the vicinity of a magnetic phase transition, and a sharp decrease below the peak signals the freezing-out of the scattering due to an onset of long-range magnetic order, respectively
\cite{17,79}. This is consistent with the magnetic measurement result.}

\subsection{XPS Result}
\begin{figure}
	\centering
		\includegraphics[width=8.5cm]{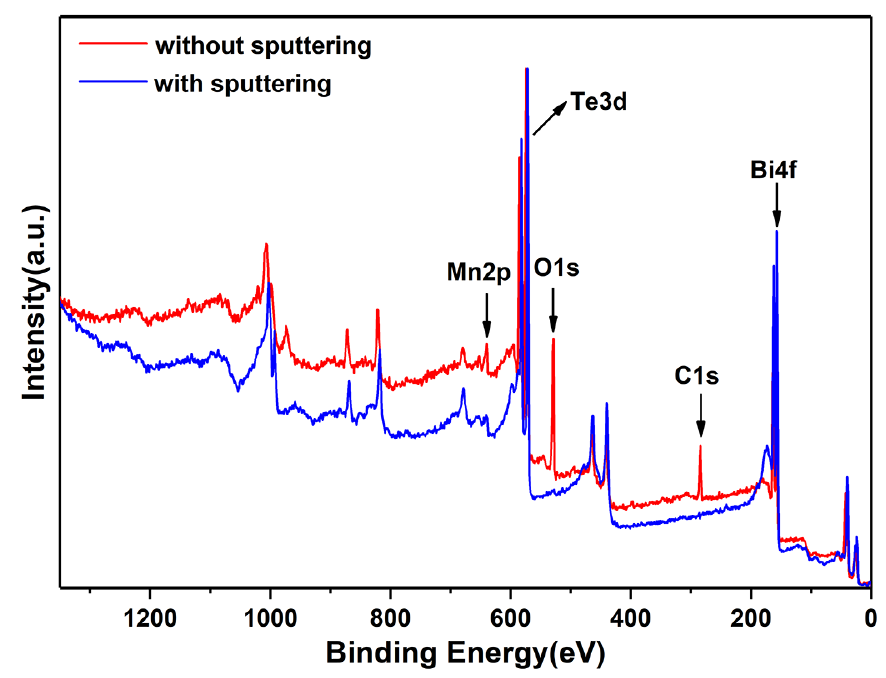}
	\caption{ Survey XPS spectra acquired from the MnBi$_2$Te$_4$ surface. After sputtering, there exists no contaminants, i.e. oxygen and carbon anymore.}
	\label{FIG:6}
\end{figure}

 The sputtering procedure is done to remove the surface contaminants of MnBi$_2$Te$_4$ (001). As illustrated by Fig. 6, compared to the initial  spectrum (above line), the spectrum  obtained from the  cleaned surface (below line) does not show surface contaminants, i.e. oxygen and carbon anymore.
\begin{figure*}
	\centering
		\includegraphics[width=14cm]{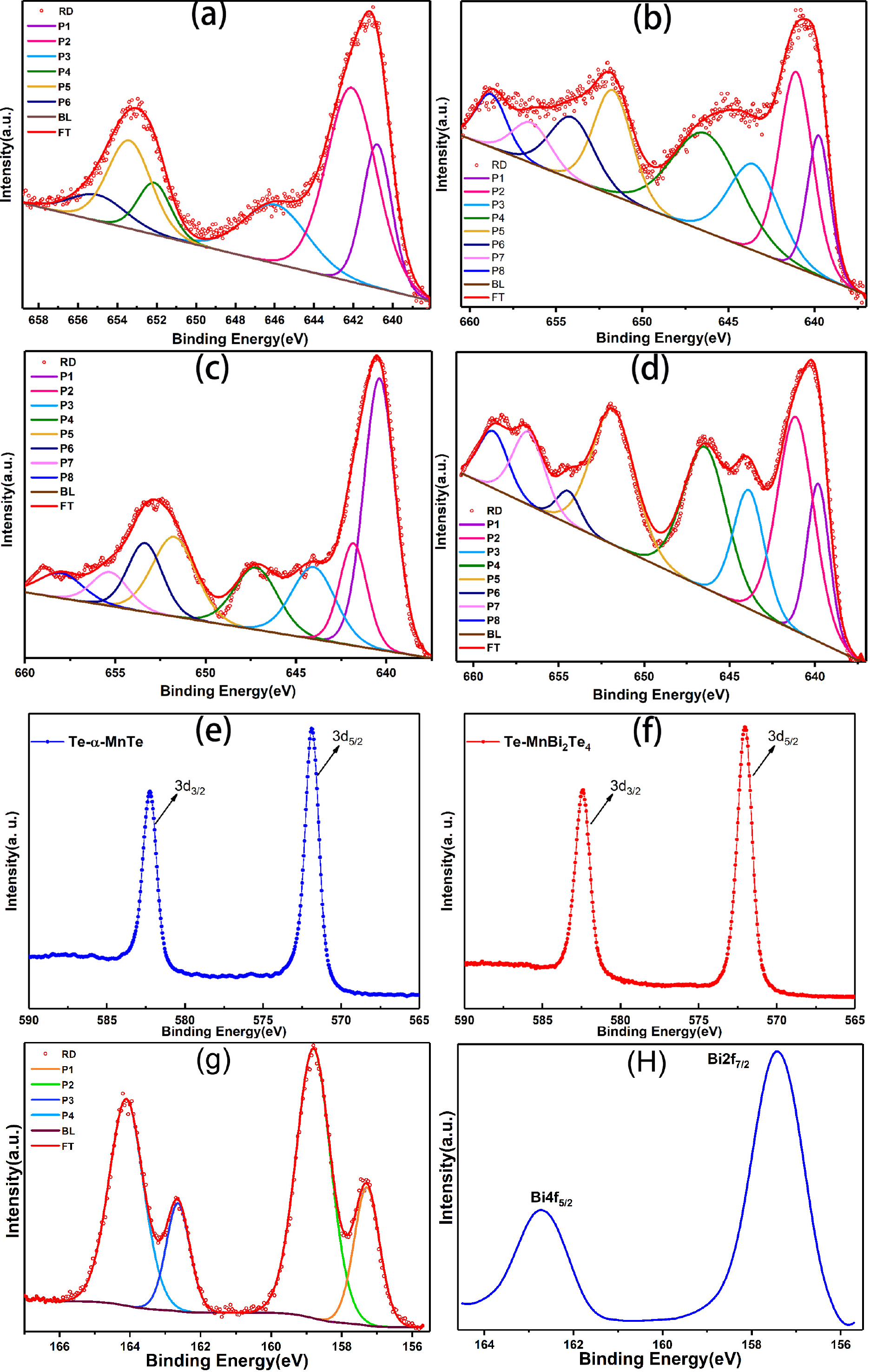}
	\caption{(a) and (b) are the XPS spectra of Mn2p core-level acquired from $\alpha$-MnTe, without and with sputtering, respectively. (c)  and (d) are the XPS spectra of Mn2p core-level acquired from the MnBi$_2$Te$_4$ surface without and with  sputtering, respectively. (e) and (f) are the XPS spectra of Te3d obtained from sputtered $\alpha$-MnTe and MnBi$_2$Te$_4$, respectively. (g) and (h) are the XPS spectra of Bi4f core-level acquired from the MnBi$_2$Te$_4$ surface without and with  sputtering, respectively. For all the panels, the peaks are numbered increasingly from right to left as Peak1 (P1), Peak 2 (P2), Peak 3 (P3), ... . Meanwhile, RD denotes the raw data; BL means the base line; FT corresponds to the fitted line.}
	\label{FIG:7}
\end{figure*}
  \renewcommand\thetable{\Roman{table}}
\begin{table}
\flushright
\setlength\aboverulesep{0pt}\setlength\belowrulesep{0pt}
\setcellgapes{3pt}\makegapedcells
\caption{ shows the fitted peaks' position of Mn2p and Te3d of $\alpha$-MnTe and MnBi$_2$Te$_4$, respectively. 'S' denotes sputtering; 'WTS' means without sputtering}
\setlength{\tabcolsep}{1.8mm}{
\begin{tabular}{c|c|cccc}
\toprule
\multirow{15}[30]{*}{Mn2P$_{3/2}$} & \multirow{7}[8]{*}{MnTe} & WTS & P1 & 640.8 & MnO and MnTe \\
\cmidrule{3-6} & & S & P1 & 639.8 & Mn subtelluride \\
\cmidrule{3-6} & & WTS & P2 & 642.1 & MnO$_2$ \\
\cmidrule{3-6} & & S & P2 & 641.1 & MnTe \\
\cmidrule{3-6} & & WTS & P3 & 646.1 & Satellite of P1 \\
\cmidrule{3-6} & & S & P3 & 643.5 & Satellite of P1 \\
\cmidrule{3-6} & & S & P4 & 646.3 & Satellite of P2 \\
\cmidrule{2-6} & \multirow{8}[10]{*}{MnBi$_2$Te$_4$} & WTS & P1 & 640.6 & MnO and MnTe \\
\cmidrule{3-6} & & S & P1 & 639.8 & Mn subtelluride \\
\cmidrule{3-6} & & WTS & P2 & 642.1 & MnO$_2$ \\
\cmidrule{3-6} & & S & P2 & 641.1 & MnTe \\
\cmidrule{3-6} & & WTS & P3 & 644.2 & Satellite of P1 \\
\cmidrule{3-6} & & S & P3 & 643.9 & Satellite of P1 \\
\cmidrule{3-6} & & WTS & P4 & 647.4 & Satellite of P2 \\
\cmidrule{3-6} & & S & P4 & 646.4 & Satellite of P2 \\
\midrule
\multirow{2}[4]{*}{Te3d$_{5/2}$} & MnTe & S & & 571.9 & Mn-Te \\
\cmidrule{2-6} & MnBi$_2$Te$_4$ & S & & 572.0 & Mn-Te \\
\midrule
\multirow{3}[3]{*}{Bi4f$_{7/2}$} & \multirow{3}[3]{*}{MnBi$_2$Te$_4$ } & WTS & P1 & 157.3 & Bi-Te \\
\cmidrule{3-6} & & WTS & P2 &158.8 & Bi$_2$O$_3$ \\
\cmidrule{3-6} & & S & &157.4 & Bi-Te \\
\bottomrule
\end{tabular}}%
\label{tab:addlabel}%
\end{table}%
Fig. 7 shows the XPS spectrum of Mn2p and Te3d. Their fitted parameters are shown in Table \uppercase\expandafter{\romannumeral1}. Deconvolution of the Mn2p spectrum of Fig. 7(a) reveals two main contributions: Peak 1 at the position of 640.8eV and Peak 2 with $E_B$=642.1 eV. Compared with the reference data \cite{24}, it is found that Peak 2 corresponds to Mn-O bond of MnO$_2$. Considering the overlap of the binding energy of MnO and MnTe\cite{25}, Peak 1 of Fig.7 (a) may be a sum of two partial contributions: MnO and MnTe. However, it is difficult to separate Peak 1 into two reasonable parts. Peak 3 shifting  to higher energies can be ascribed to the satellite of Peak 1\cite{25,26,27,28,29}. The satellites are also called shakeup satellites. It locates at the higher-binding-energy side of the relevant core-lines of transition metal (TM) or rare earth (RE) ions in the selected TM or RE compounds \cite{30}. It also concerns on particular binary Mn oxides, especially MnO \cite{29}. Such satellite structure in Mn 2p XPS spectrum originates from charge transfer between outer electron shell of ligand and an unfilled 3d shell of Mn during creation of core-hole in the photoelectron process \cite{27,28,29}. For Fig.7 (c), Peak 1 at the position of 639.8eV corresponds to +1 charge state of Mn, which may come from nanoclusters of manganese subtelluride (possibly Mn$_2$Te) induced by $Ar^+$ ion sputtering\cite{25}.The above effect is strictly analogous to the reduction of Mn oxides during $Ar^+$ ion bombardment\cite{31}. Peak 3 with the binding energy of 643.5 eV is supposed to be the satellite of Peak 1. Peak 2 at the position of 642.1eV belongs to Mn-O bonds of MnO$_2$. Peak 4 is the satellite of Peak 2. For (b) and (d) panels of Fig.6, the removal of contaminants from the surface leads to disappearance of the native oxides contributions to the Mn2p spectrum, i.e. those of MnO$_2$ and MnO. Therefore, Peak 2 of Fig. 7(d), with $E_B$=641.1 eV, can be ascribed to the divalent Mn-Te bond of MnBi$_2$Te$_4$. Peak 4 of Fig. 7(d) exhibiting a shift from Peak 2 with the value of $\Delta$E=$E_{p4}$-$E_{p2}$= 5.3 eV (as shown by Table  \uppercase\expandafter{\romannumeral1}) is the satellite of Peak 2. Peak 1 with $E_B$ = 639.8 eV may be induced by the $Ar^+$ ion bombardment during the surface preparation. It is the +1 charge state of Mn and can be ascribed to the Mn–Te bond in the nanoclusters of manganese subtelluride (possibly Mn$_2$Te), created in the surface and subsurface region penetrated by $Ar^+$ ions \cite{32}. Peaks 3 is the corresponding satellite of Peak 1. In addition, rising intensity of the Mn2p satellite for divalent Mn (bound to chalcogen) with atomic number of ligand (from MnO to MnBi$_2$Te$_4$) could be observed too. As shown by Fig. 7 (e) and (f) that the peaks’ positions of Te3d of $\alpha$-MnTe almost overlap with those of MnBi$_2$Te$_4$, which means that the chemical states of Te both from $\alpha$-MnTe and  MnBi$_2$Te$_4$ are the same.  {Fig. 7 (g) and (h) are the XPS spectra of Bi4f acquired from the MnBi$_2$Te$_4$ surface without and with  sputtering, respectively. As shown by Fig.7(g) that peak 1 locating at the position of 157.3eV corresponds to Bi-Te bond, while  peak 2 locating at the position of 158.8eV corresponds to Bi-O bond of Bi$_2$O$_3$\cite{77, 78}. As shown by Fig.7(h) that, after being sputtered, there exists only one kind of bond i. e. Bi-Te bond with the binding energy of 157.4eV. This tells us again that the contaminants have been removed completely by the sputtering procedure.} Comparison of the fitted Mn2p and Te3d parameters of the MnBi$_2$Te$_4$ and $\alpha$-MnTe spectral components gathered in Table \uppercase\expandafter{\romannumeral1}, it shows a very good consistency between them (considering the accuracy of the method, i.e. $\Delta$E = 0.1 eV).  Then we can get the conclusion that, Mn and Te of MnBi$_2$Te$_4$   share the same chemical state with those of  $\alpha$-MnTe, which supports our hypothesis that the growth mode of MnBi$_2$Te$_4$ is the inserting-layer growth mode.
\section{Conclusion}\par
Single crystals of antiferromagnetic topological insulator MnBi$_2$Te$_4$ have been grown by flux method. The magnetic property of MnBi$_2$Te$_4$ is investigated, which shows good antiferromagnetic character with Neel temperature of 24.5K and a spin-flop transition at H$\thickapprox$ 35000 Oe, 1.8K. The possible inserting-layer growth mode of MnBi$_2$Te$_4$ is proposed, which is supported by the XPS and XRD result of $\alpha$-MnTe and MnBi$_2$Te$_4$. The proposed crystal growth mechanism may help us to grow the  other members of MB$_2$T$_4$-family materials, which have rich topological quantum states.\par
The XPS measurement combing with the $Ar^+$ ion sputtering is done to investigate chemical states of MnBi$_2$Te$_4$ and $\alpha$-MnTe. Binding energies of MnBi$_2$Te$_4$-related contributions to the Mn2p and Te3d spectra agree well with those of $\alpha$-MnTe. \par
\section{Acknowledgements}\par
This work was supported by the Natural Science Foundation of Shandong Province, China (Nos. ZR2016AQ08, ZR2019MA036) and National Natural Science Foundation of China (Nos. 11804194).
\bibliographystyle{unsrt}

\bibliography{ref}

\end{document}